\documentclass[aps,prl,reprint,twocolumn,superscriptaddress, showpacs,10pt]{revtex4}

\usepackage{times}
\usepackage{color,graphicx}
\usepackage{subfig,array}
\usepackage{amsthm,amssymb,amsmath}

\usepackage{amscd}

\newcommand\ket[1]{\ensuremath{|#1\rangle}}
\newcommand\bra[1]{\ensuremath{\langle#1|}}

\newcommand\tr{\mathop{\rm tr}\nolimits}

\def\a{\mathbf{a}}

\newtheorem{conjecture}{Conjecture}
\newtheorem{definition}{Definition}
\newtheorem{lemma}{Lemma}
\newtheorem{theorem}{Theorem}
\newtheorem{corollary}{Corollary}
\newtheorem{remark}{Remark}
\newtheorem{proposition}{Proposition}
\newtheorem{observation}{Observation}

\newtheorem{question}{Question}
\newtheorem{example}{Example}

\def\SL{{\mbox{\rm SL}}}
\def\SU{{\mbox{\rm SU}}}

\def\Dbar{\leavevmode\lower.6ex\hbox to 0pt
{\hskip-.23ex\accent"16\hss}D}

\def\bcj{\begin{conjecture}}
\def\ecj{\end{conjecture}}
\def\bcr{\begin{corollary}}
\def\ecr{\end{corollary}}
\def\bd{\begin{definition}}
\def\ed{\end{definition}}
\def\bea{\begin{eqnarray}}
\def\eea{\end{eqnarray}}
\def\bem{\begin{enumerate}}
\def\eem{\end{enumerate}}
\def\bex{\begin{example}}
\def\eex{\end{example}}
\def\bim{\begin{itemize}}
\def\eim{\end{itemize}}
\def\bl{\begin{lemma}}
\def\el{\end{lemma}}
\def\bpf{\begin{proof}}
\def\epf{\end{proof}}
\def\bpp{\begin{proposition}}
\def\epp{\end{proposition}}
\def\bqu{\begin{question}}
\def\equ{\end{question}}
\def\br{\begin{remark}}
\def\er{\end{remark}}
\def\bt{\begin{theorem}}
\def\et{\end{theorem}}

\def\lin{\mathop{\rm span}}
\def\we{\wedge}

\def\bC{{\mathbb{C}}}
\def\sue{\subseteq}

\def\hdet{{\mbox{\rm Det}}}
\def\Un{{\mbox{\rm U}}}

\newcommand{\nc}{\newcommand}

\nc{\cA}{{\cal A}} \nc{\cB}{{\cal B}} \nc{\cC}{{\cal C}}
\nc{\cD}{{\cal D}} \nc{\cE}{{\cal E}} \nc{\cF}{{\cal F}}
\nc{\cG}{{\cal G}} \nc{\cH}{{\cal H}} \nc{\cI}{{\cal I}}
\nc{\cJ}{{\cal J}} \nc{\cK}{{\cal K}} \nc{\cL}{{\cal L}}
\nc{\cM}{{\cal M}} \nc{\cN}{{\cal N}} \nc{\cO}{{\cal O}}
\nc{\cP}{{\cal P}} \nc{\cQ}{{\cal Q}} \nc{\cR}{{\cal R}}
\nc{\cS}{{\cal S}} \nc{\cT}{{\cal T}} \nc{\cU}{{\cal U}}
\nc{\cV}{{\cal V}} \nc{\cW}{{\cal W}} \nc{\cX}{{\cal X}}
\nc{\cZ}{{\cal Z}}

\newcommand{\braket}[2]{\langle#1|#2\rangle}

\def\sue{\subseteq}

\def\we{\wedge}
\def\diag{\mathop{\rm diag}}

\def\a{\alpha}

\def\i{\iota}

\def\s{\sigma}

\begin{document}


\title{Four-qubit pure states as fermionic states}

\author{Lin Chen}%
\affiliation{Department of Pure Mathematics, University of Waterloo, Waterloo, Ontario, Canada}
\affiliation{Institute for Quantum Computing, University of Waterloo,
  Waterloo, Ontario, Canada}%
\author{Dragomir {\v{Z} {\Dbar}}okovi{\'c}}%
\affiliation{Department of Pure Mathematics, University of Waterloo, Waterloo, Ontario, Canada}
\affiliation{Institute for Quantum Computing, University of Waterloo,
  Waterloo, Ontario, Canada}%
\author{Markus Grassl}%
\affiliation{Centre for Quantum Technologies, National University of Singapore, Singapore}
\author{Bei Zeng}%
\affiliation{Institute for Quantum Computing, University of Waterloo,
  Waterloo, Ontario, Canada}%
\affiliation{Department of Mathematics \& Statistics, University of
  Guelph, Guelph, Ontario, Canada}%

\begin{abstract}
The embedding of the $n$-qubit space into the $n$-fermion space with
$2n$ modes is a widely used method in studying various aspects of
these systems. This simple mapping raises a crucial question: does the
embedding preserve the entanglement structure? It is known that the
answer is affirmative for $n=2$ and $n=3$. That is, under either local
unitary (LU) operations or with respect to stochastic local operations
and classical communication (SLOCC), there is a one-to-one
correspondence between the $2$- (or $3$)-qubit orbits and the $2$- (or
$3$)-fermion orbits with $4$ (or $6$) modes. However these results do
not generalize as the mapping from the $n$-qubit orbits to the
$n$-fermion orbits with $2n$ modes is no longer surjective for
$n>3$. Here we consider the case of $n=4$. We show that surprisingly,
the orbit mapping from qubits to fermions remains injective under
SLOCC, and a similar result holds under LU for generic orbits.  As a
byproduct, we obtain a complete answer to the problem of SLOCC
equivalence of pure $4$-qubit states.
\end{abstract}

\date{\today}


\maketitle

It is well-known that any $n$-qubit pure state $\ket{\psi}$ can be
`viewed' as an $n$-fermion state with $2n$ modes. The underlying
reason is that one can `pair' those $2n$ modes to obtain $n$ pairs,
and then allow only one mode of each pair to be `occupied' by a
fermion (i.\,e. `single occupancy'). This simple mapping has been
used widely as a technique to study various aspects of qubits and fermionic
systems, such as the general relationship between fermionic systems
and spin systems \cite{VC05}, the QMA-completeness of the
$N$-representability problem \cite{LCV07}, the black hole/qubit
correspondence \cite{BDL12}, and ground state properties of fermionic
systems with local Hamiltonians \cite{OCZ+11}.

Despite the success of this simple embedding from the $n$-qubit space
into an $n$-fermion space with $2n$ modes, the crucial question
whether the mapping preserves the entanglement structure remains
unclear.  At first glance, this seems quite implausible as after the
embedding the qubit local group is only a `small' subgroup of the
fermionic local group.  However, it was shown in \cite{SLM01,SCK+01}
that the LU orbits of the $2$-fermion system with $4$ modes are in a
one-to-one correspondence with the $2$-qubit LU orbits, based on the
fermionic version of Schmidt decomposition, and the result also holds
when considering SLOCC orbits.  The $n=3$ SLOCC case was discussed in
\cite{LV08}, and surprisingly, the one-to-one correspondence of orbits
stays intact.  In fact, the mathematical problem has been studied in
multilinear algebra and matrix analysis for many years
\cite{Sch31,Ehr99}. For the $n=3$ LU case, related studies in the $N$-representability
community for the $3$-fermion system gave some hints
\cite{BD72,Rus07,Kly06}, and recently it is shown that
the mysterious one-to-one relationship between orbits
remains \cite{CDGZ13}.

A simple dimension counting shows that the one-to-one relationship
between orbits does not generalize for $n>3$ \cite{CCD+13}.  There are
indeed more fermionic orbits than qubit orbits.  However, it is
natural to ask whether the fermionic local groups can mix any two
locally inequivalent qubit states.  Yet this seems to be tough
question, even for the $n=4$ case, for several reasons: the $4$-qubit
orbits under SLOCC received various controversial treatments,
manifesting that this by itself is not a simple problem
\cite{VDDV02,BDD+10,LLSS07,CD06,Wal04,LLS+07,CW07,LLHL08,AG10}; under
SLOCC there are infinitely many orbits for $n=4$, while for $n=3$
there are only finitely many orbits; unlike the LU case, knowing only
the invariants does not solve the SLOCC classification problem; unlike
the $n=3$ case, only little is known for LU invariants in the $n=4$
case.

In this work, we handle the $n=4$ case.  We show that surprisingly,
two inequivalent $4$-qubit states under the qubit SLOCC group
$\SL_2^{\times 4}\rtimes S_4$ (including qubit permutations) remain
inequivalent under the fermionic SLOCC group $\SL_8$. We ignore the
constant factor introduced from replacing the `true' SLOCC group by
matrices with determinant one, as the corresponding success
probability of the SLOCC protocol is not important for our
discussion.  Our proof relies on the celebrated theorem of Kostant
and Rallis that for an (infinitesimal) symmetric space, any vector
admits a unique Jordan decomposition into a semisimple part and a
nilpotent part which commute \cite{KR71}.  We examine separately
the semisimple, nilpotent, and mixed orbits. As a byproduct, we
complete the SLOCC classification for $4$-qubit states given in
\cite{VDDV02,CD06}.  Furthermore, we show that generically, two
inequivalent $4$-qubit states under the LU group $\SU_2^{\times
  4}\rtimes S_4$ remain inequivalent under the fermionic LU group
$\SU_8$.

\textit{The setting}---We denote by $V$ an $8$-dimensional complex
Hilbert space, for which we fix an orthonormal basis
$\{\ket{i}\colon i=1,\ldots,8\}$ and the orthogonal decomposition
$V=V_1\oplus V_2\oplus V_3\oplus V_4$ into four $2$-dimensional
subspaces $V_i:=\lin\{\ket{2i-1},\ket{2i}\}$. The exterior power
$\we^4(V)$ is the Hilbert space of a fermionic system consisting of
$4$ fermions with $8$ modes. The $4$-vectors
$e_{ijkl}:=\ket{i\we j\we k\we l}$, $1\le i<j<k<l\le8$, form an
orthonormal basis of $\we^4(V)$. We shall view the tensor product
${\cal H}:=V_1\otimes V_2\otimes V_3\otimes V_4$ as the Hilbert
space of $4$ qubits. We identify this tensor product with the
subspace $W:=V_1\we V_2\we V_3\we V_4$ of $\we^4(V)$ via the
isometric embedding
 \bea \label{eq:embedding}
\ket{ijkl} \mapsto e_{1+i,3+j,5+k,7+l}, \quad
i,j,k,l\in\{0,1\}.
 \eea

\begin{figure}[h]
\setlength{\unitlength}{0.14in} 
\centering 
\begin{picture}(24,7) 
\put(0,4){ \fbox{
 $\begin{array}{l}
 \; \\
 f_1\big\uparrow:=1\\
 \; \\
 f_1\big\downarrow:=2 \\
 \;\\
 \end{array}$
 }}
 \put(6,4){ \fbox{
 $\begin{array}{l}
 \; \\
 f_2\big\uparrow:=3\\
 \; \\
 f_2\big\downarrow:=4 \\
 \;\\
 \end{array}$
 }}
  \put(12,4){ \fbox{
 $\begin{array}{l}
 \; \\
 f_3\big\uparrow:=5\\
 \; \\
 f_3\big\downarrow:=6 \\
 \;\\
 \end{array}$
 }}
  \put(18,4){ \fbox{
 $\begin{array}{l}
 \; \\
 f_4\big\uparrow:=7\\
 \; \\
 f_4\big\downarrow:=8 \\
 \;\\
 \end{array}$
 }}
\put(2.5,0){$V_1$}
\put(8.5,0){$V_2$}
\put(14.5,0){$V_3$}
\put(20.5,0){$V_4$}
\end{picture}
\caption{A physical picture of the embedding of
Eq.~\eqref{eq:embedding}. Each subspace $V_i$
labels a specially localized site $i$ with spacial mode $f_i$,
and each $\uparrow$ ($\downarrow$) denotes a spin up (down) state.}
\label{fig:SOVfig}
\end{figure}
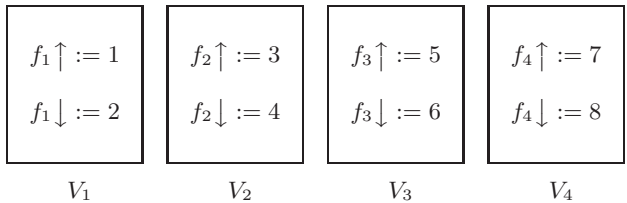

Physically, under this embedding, any $4$-qubit state can be viewed as a fermionic single occupancy vector (SOV). One can imagine each of the subspaces $V_i$ as a localized site (or an atomic orbit) with two electron spin states, as illustrated in Fig.~\ref{fig:SOVfig}. Then for an SOV, each $V_i$ can only be
`occupied' by a single fermion. The SOV space is then identified
with ${\cal H}$.

There are natural groups that act on these spaces. On ${\cal H}$ we
have the action of the LU group $\SU:=\SU(2,\bC)^{\times4}$ and the
SLOCC group $\SL:=\SL(2,\bC)^{\times4}$. We can extend these groups by
including the symmetric group $S_4$ which permutes the four qubits,
and the extended groups are the semidirect product $\SU\rtimes S_4$
and $\SL\rtimes S_4$. On the space $\we^4(V)$ we have the natural
action of the LU group $\SU_8:=\SU(8)$ and the SLOCC group
$\SL_8:=\SL(8,\bC)$.

Notice that under the embedding the qubit local group $\SU\rtimes S_4$
(or $\SL\rtimes S_4$) is a subgroup of the fermionic local group
$\SU_8$ (or $\SL_8$).  Despite these larger local groups in
the fermionic case, for a similar embedding from $2$- or $3$-qubit
states to fermionic SOVs, it is found that there is a one-to-one
correspondence between the qubit orbits and fermionic orbits under
both SLOCC and LU (including qubit permutations)
\cite{SLM01,LV08,CDGZ13}.  However, these results will not directly
generalize to more than $3$ qubits, as there exist fermionic orbits
which do not meet the SOV space \cite{CCD+13}.

Our study is motivated by the following intriguing question:
\textit{if two pure $4$-qubit states are inequivalent under the qubit
  local group enlarged by permutations, $\SU\rtimes S_4$
(or $\SL\rtimes S_4$), do they
  remain inequivalent under the fermionic group $\SU_8$ (or $\SL_8$) after the
  embedding into the fermionic system?}

Notice that a similar question was asked in \cite{VL09}, without
considering qubit permutations. In that case simple counter-examples
can be found, where qubit states which are not equivalent under
$\SL_{n}^{\times 2}$ become equivalent under the fermionic
$\SL_{2n}$. However, the question becomes much harder when
qubit permutations are considered.

Surprisingly, we shall show that the answer to our question
is affirmative for the SLOCC case, and almost always affirmative
for the LU case. We shall first discuss the relatively simpler
case of SLOCC.

\textit{The SLOCC case}---For convenience, let us
introduce the following notation. For any $(\SL\rtimes S_4)$-orbit
${\cal O}\sue {\cal H}$ we shall denote by $\tilde{{\cal O}}$ the
unique $\SL_8$-orbit in $\we^4(V)$ which contains ${\cal O}$.
Our main result is then given by the following theorem.

\bt
\label{thm:main}
If ${\cal O}_1\ne{\cal O}_2$ are two $(\SL\rtimes S_4)$-orbits in ${\cal H}$,
then $\tilde{{\cal O}}_1\ne\tilde{{\cal O}}_2$.
\et

Note that the $\SL_8$-orbits in $\we^4(V)$ were classified in
\cite{An81}, and the $\SL\rtimes S_4$-orbits were first classified in
\cite{VDDV02}, giving the well-known results of nine families.  There
are subsequent treatments in
\cite{BDD+10,LLSS07,CD06,Wal04,LLS+07,CW07, LLHL08,AG10}, and we shall
use the nine families obtained in \cite{CD06} which provides some
corrections to \cite{VDDV02}.

\textit{The case of closed orbits}---To prove Theorem~\ref{thm:main},
we first consider the closed orbits. This is natural since almost all
orbits are closed.  We shall show that in this case the answer to our
question is affirmative, as given by the following

\begin{observation}
\label{pp:ClosedOrbits}
Let $\cO_1$ and $\cO_2$ be two
different closed $\SL\rtimes S_4$-orbits. Then, after the embedding
$\cH\to\we^4(V)$ into the fermionic system, the enlarged orbits
$\tilde{\cO}_1$ and
$\tilde{\cO}_2$ are also closed and different.
\end{observation}

In order to show this, we shall use some facts from the
theory of invariants (see e.\,g. \cite{PV94}).

Let $\cA$ be the algebra of complex polynomial functions on $\we^4(V)$
which are invariant under the action of $\SL_8$. Let $\we^4(V)/\SL_8$
denote the affine variety attached to the algebra $\cA$. It is known
that $\cA$ is isomorphic to the polynomial algebra over $\bC$ in seven
variables (see e.\,g. \cite[Sec.~1.3]{Katanova92}). Consequently, as an
affine variety, $\we^4(V)/\SL_8$ is isomorphic to the affine space
$\bC^7$.  The importance of this variety is that it parametrizes the
closed $\SL_8$-orbits in $\we^4(V)$. More precisely, each closed orbit
is represented by a point in the variety, each point in the variety
corresponds to some closed orbit, and different points represent
different closed orbits.

Similarly, let $\cB$ be the algebra of complex polynomial functions
on $\cH$ which are invariant under the action of $\SL\rtimes S_4$.
It is also known \cite{CD06} that $\cB$ is isomorphic to the
polynomial algebra over $\bC$ in four variables. Consequently, the
affine variety $\cH/(\SL\rtimes S_4)$ attached to $\cB$ is
isomorphic to the affine space $\bC^4$. This variety parametrizes
the closed $\SL\rtimes S_4$-orbits in $\cH$.

Note that, due to our embedding \eqref{eq:embedding}, the group
$\SL\rtimes S_4$ is a subgroup of $\SL_8$. For any polynomial
function $f\colon\we^4(V)\to\bC$ we shall denote by $f'$ its
restriction to the subspace $\cH$. Note that if $f\in\cA$ then
$f'\in\cB$. This restriction map $\cA\to\cB$ is a homomorphism of
algebras. By working with explicit generators of the algebras $\cA$
and $\cB$, for details see Appendix A, we shall prove that
this homomorphism is onto and consequently the corresponding morphism
of varieties
 \bea \label{eq:morphism}
\Phi: \bC^4 \cong \otimes_i V_i/(\SL\rtimes S_4) \to \we^4(V)/\SL_8
\cong \bC^7
 \eea
is injective.

We now claim that if $\psi\in\cO$, where $\cO$ is a closed
$(\SL\rtimes S_4)$-orbit in $\cH$, then the orbit
$\tilde{\cO}=\SL_8\cdot\psi$ is also closed.
To show this, we adopt the method in \cite{An81},
to use the locally symmetric space
$(\mathfrak{e}_7,\mathfrak{sl}_8)$ with Cartan decomposition
\begin{equation}
\label{eq:E_7}
\mathfrak{e}_7=\mathfrak{sl}_8\oplus \mathfrak{p}.
\end{equation}
This means that the linear map $\theta$ in \cite{An81}, on the
exceptional complex simple Lie algebra $\mathfrak{e}_7$, having
$\mathfrak{sl}_8$ and  $\mathfrak{p}$ as its $+1$ and $-1$
eigenspaces, respectively, is an involutory automorphism of
$\mathfrak{e}_7$.  Moreover, the subspace $\mathfrak{p}$ can be
identified with $\we^4(V)$ as an $\SL_8$-module. The closed
$\SL_8$-orbits in $\we^4(V)$ are precisely those that meet the Cartan
subspace $\mathfrak{c}\sue\mathfrak{p}$ ($\mathfrak{c}$ has dimension
seven and is unique up to the action of $\SL_8$). In
\cite[Sect.~3.1]{An81}, the following basis elements of $\mathfrak{c}$
are given:
\begin{eqnarray}
p_1 :&=& e_{1234}+e_{5678},\quad p_2 := e_{1357}+e_{6824},\nonumber\\
p_3 :&=& e_{1562}+e_{8437},\quad p_4 := e_{1683}+e_{4752},\nonumber\\
p_5 :&=& e_{1845}+e_{7263},\quad p_6 := e_{1476}+e_{2385},\nonumber\\
p_7 :&=& e_{1728}+e_{3546}.
\end{eqnarray}

The closed $\SL\rtimes S_4$-orbits are those that belong to the family
1 in \cite[Table 7]{CD06}. These are precisely the orbits that meet
the $4$-dimensional subspace $\mathfrak{a}$ spanned by the vectors
\begin{eqnarray}
\ket{0000}+\ket{1111},\quad &&\ket{0011}+\ket{1100},\nonumber\\
\ket{0101}+\ket{1010},\quad &&\ket{0110}+\ket{1001}.
\end{eqnarray}
After the embedding into the fermionic space, these vectors become
$p_2$, $p_4$, $p_5$, and $-p_6$, respectively. Since this subspace is
contained in $\mathfrak{c}$, our claim follows.

We summarize this in the following diagram,
\begin{eqnarray}
\begin{array}{rccccrc}
    \mathfrak{e}_7\ &=\ &\mathfrak{sl}_8&\oplus&\we^4(V)&\supseteq &\mathfrak{c} \\[0.5ex]
    \big\uparrow\ &&\ \big\uparrow&&\big\uparrow&&\big\uparrow \\[0.5ex]
    \mathfrak{so}_8\ &=&\ \mathfrak{sl}_2^{\times 4}&\oplus&
\bigotimes_{i=1}^4 V_i &\supseteq &\mathfrak{a} \\
\end{array}
\end{eqnarray}
where $\mathfrak{e}_7$, $\mathfrak{so}_8$, 
$\mathfrak{sl}_2^{\times4}$ are the
corresponding Lie algebras, and the arrows $\uparrow$ indicate the corresponding embeddings. 
In Appendix B we provide more details about these embeddings by using the Dynkin diagrams.

We can now show Observation~\ref{pp:ClosedOrbits}. We have just
shown that the orbits $\tilde{\cO}_i$ are closed and that
$\Phi(\cO_i)=\tilde{\cO}_i$ for $i=1,2$. Since $\Phi$ is one-to-one
and $\cO_1\ne\cO_2$ we have $\Phi(\cO_1)\ne\Phi(\cO_2)$, i.\,e.,
$\tilde{\cO}_1\ne\tilde{\cO}_2$.

\textit{The case of nilpotent orbits}---An $\SL$- or $(\SL\rtimes
S_4)$-orbit $\cO\sue \cH$ is called {\em nilpotent} if its closure
contains the zero vector. One defines similarly the nilpotent
$\SL_8$-orbits in $\we^4(V)$.  In the previous section we have shown
that if $\cO$ is closed then $\tilde{\cO}$ is closed as well. On the
other hand, it is immediate from the definition, that if $\cO$ is
nilpotent then $\tilde{\cO}$ is nilpotent, too. Among the nilpotent
orbits, only the trivial orbit $\{0\}$ is closed.

There are exactly $9$ nilpotent $(\SL\rtimes S_4)$-orbits (including
the trivial orbit $\{0\}$). The representatives of these $9$ orbits
can be obtained from \cite[Table 7]{CD06} by setting the parameters
$a, b, c, d$ (if any) to $0$. That table classifies the $(\SL\rtimes
S_4)$-orbits in $\cH$ into $9$ families depending on the complex
parameters $a$, $b$, $c$, $d$. They are numbered by integers 1, 2, 3,
6, 9, 10, 12, 14, and 16. The last three families consist of a single
orbit, which is nilpotent. Two states belonging to different families
do not belong to the same $(\SL\rtimes S_4)$-orbit \cite[Theorem
  3.6]{CD06}. In particular, the $9$ nilpotent orbits are pairwise
distinct.

The main result of this section is the following observation.

\begin{observation}
 \label{pp:SOV-nilp}
If $\cO_1$ and $\cO_2$ are nilpotent $(\SL\rtimes S_4)$-orbits and
$\cO_1\ne\cO_2$, then $\tilde{\cO}_1\ne\tilde{\cO}_2$.
\end{observation}

To show this observation, clearly we may assume that both $\cO_1$ and
$\cO_2$ are non-zero orbits. We shall say that an $\SL_8$-orbit in
$\we^4(V)$ is an {\em SOV orbit} if it meets $\cH$. Trivially, if
$\cO$ is an $(\SL\rtimes S_4)$-orbit, then $\tilde{\cO}$ is an SOV
orbit. However, there exist $\SL_8$-orbits in $\we^4(V)$ which are not
SOV \cite{CCD+13}. To show the observation, it suffices to show that
there are at least $8$ non-zero nilpotent $\SL_8$-orbits in $\we^4(V)$
which are SOV.

There are exactly $94$ non-zero nilpotent $\SL_8$-orbits in
$\we^4(V)$, see \cite[Table 2]{An81} (as well as also \cite[Table
  XI]{Dj88} for an independent derivation carried out in a different
context).  Both enumerations make use of the decomposition
\eqref{eq:E_7}, where $\mathfrak{p}=\we^4(V)$. The nilpotent
$\SL_8$-orbits in $\we^4(V)$ are then classified by constructing
representatives of the so-called {\em normal
  $\mathfrak{sl}_2$-triples}. These are non-zero triples $(H,E,F)$
with $H\in\mathfrak{sl}_8$ and $E,F\in\we^4(V)$ such that
\begin{alignat}{5}
[H,E]=2E,\quad [H,F]=-2F,\quad [E,F]=H.\label{eq:SL2_triple}
\end{alignat}
In \cite[Table 2]{An81}, the elements $H\in\mathfrak{sl}_8$ are
given as diagonal matrices, but the elements $E$ and $F$ were not
computed. They can be computed by using Eqs. \eqref{eq:SL2_triple}.

Our computations show that the orbits 1, 2, 5, 6, 9, 20, 44, and 50
in~\cite[Table 2]{An81} are SOV orbits. The results are summarized in
Table \ref{tab:nilpotent_orbits}, where the first column gives the
label of the orbit in~\cite[Table 2]{An81}, and the second column
gives the label of the family from \cite[Table 7]{CD06} whose
nilpotent orbit (obtained by setting $a=b=c=d=0$) is contained in the
orbit given in the first column. The third and fourth columns list the
corresponding elements $H$ and $E$.
\begin{table}[h!]
\[\def\half{\frac{1}{2}}
\def\arraystretch{1.1}
\begin{array}{c|c|c|l}
\text{No.} &\text{No.} & H=\diag(\lambda_j)
  & \text{nilpotent elements $E$} \\
\text{\cite{An81}} &\text{\cite{CD06}} &
  & \\
\hline
 1 & 2 & \half(1, 1, 1, 1,
       & e_{1234} \\
   & &-1, -1, -1, -1) & \\
\hline
 2 & 3 & (1, 1, 0, 0,
       & i(e_{1235} + e_{1246}) \\
   & & 0, 0, -1, -1)&  \\
\hline
 5 & 10 & \half(3,1,1,1,
       & i(e_{1347} + e_{1235}+ e_{1246})  \\
   & &-1,-1,-1,-3) &   \\
\hline
 6 & 6 & (1, 1, 1, 1,
       & i(e_{1347} + e_{1235}+ e_{2348} + e_{1246})  \\
   & & -1, -1, -1, -1)&  \\
\hline
 9 & 16 & (2, 0, 0, 0,
       & \sqrt{2}(e_{1234} + ie_{1567}) \\
   & & 0, 0, 0, -2)&  \\
\hline 20 & 9 & (2, 2, 1, 1,
       & i\sqrt{3}(e_{1347}+e_{2348}) + 2e_{1256} \\
   & & -1, -1, -2, -2)&  \\
\hline 44 & 14 & (2, 2, 2, 0,
       & e_{1357} + 2e_{2358} + e_{1256} \\
   & & 0, -2, -2, -2)& + i\sqrt{3}(e_{1347} - e_{1246})  \\
\hline 50 & 12 & (4, 2, 2, 2,
       & \sqrt{6}(e_{1357} + e_{1467} + e_{1256})  \\
   & &-2, -2, -2, -4) & + i\sqrt{10}e_{2348} \\
\end{array}
\]
\caption{Nilpotent SOV orbits}
\label{tab:nilpotent_orbits}
\end{table}

Note that the choice of the elements $E$ and $F$ is not unique. In
addition to (\ref{eq:SL2_triple}), we have imposed the condition that
\begin{equation}
F=\sigma E,
\end{equation}
where $\sigma$ is another involutory automorphism of the Lie algebra
$\mathfrak{e}_7$ which commutes with $\theta$. The action of $\sigma$
on $\mathfrak{sl}_8$ is given by $\sigma(X)=-X^T$, where $T$ denotes
the transposition, and on $\we^4(V)$ it is specified by the images of
the basis elements
\begin{equation}
\sigma\colon e_{ijkl}\mapsto e_{9-l,9-k,9-j,9-i}.
\end{equation}
Note that $\sigma H=-H$ because each $H$ is a diagonal matrix.

\textit{The general case}---To finish the proof of Theorem
\ref{thm:main}, we shall use some results from Kostant and Rallis
\cite{KR71}.  Any $\psi\in\we^4(V)$ admits a unique {\em Jordan
  decomposition} $\psi=\psi_s+\psi_n$, i.\,e., such that $\psi_s$ is
semisimple (as an element of the Lie algebra $\mathfrak{e}_7$ ),
$\psi_n$ is nilpotent, and they commute (that is, their Lie bracket
vanishes, $[\psi_s,\psi_n]=0$). An element $\psi\in\we^4(V)$ is
semisimple if and only if the orbit $\SL_8\cdot\psi$ is closed, which
is equivalent to the condition that this orbit meets the Cartan
subspace ${\mathfrak c}$ \cite[Chapter III, Theorem 4.19]{Helg84}.

Moreover, if $g\in\SL_8$ and $\psi\in\we^4(V)$ then
$g\cdot\psi=g\cdot\psi_s+g\cdot\psi_n$ is the Jordan decomposition
of $g\cdot\psi$, i.\,e., we have $(g\cdot\psi)_s=g\cdot\psi_s$ and
$(g\cdot\psi)_n=g\cdot\psi_n$.

Let $\psi\in\we^4(V)$ be one of the elements corresponding to the
states $|\psi\rangle$ listed in~\cite[Table 7]{CD06}.  Denote by
$\psi''$ the nilpotent element obtained from $\psi$ by setting
$a=b=c=d=0$ and set $\psi':=\psi-\psi''$.  We then observe that
$\psi=\psi'+\psi''$ is the Jordan decomposition of $\psi$, i.\,e., we
have $\psi_s=\psi'$ and $\psi_n=\psi''$. Indeed, we know that $\psi''$
is nilpotent and it is easy to verify that $\psi'\in{\mathfrak c}$,
and so $\psi'$ is semisimple. Furthermore, one can verify that $\psi'$
and $\psi''$ commute for all values of the parameters $a,b,c,d$. By
the uniqueness of the Jordan decomposition, we must have
$\psi_s=\psi'$ and $\psi_n=\psi''$.

To prove Theorem~\ref{thm:main}, by \cite[Theorem 3.6]{CD06} we may
assume that $\cO_1=(\SL\rtimes S_4)\cdot\phi$ and $\cO_2=(\SL\rtimes
S_4)\cdot\psi$, where $\phi$ and $\psi$ are some pure $4$-qubit states
listed in \cite[Table 7]{CD06} (possibly in different families).
Assume that $\tilde{\cO}_1=\tilde{\cO}_2$. Then $\psi=g\cdot\phi$ for
some $g\in\SL_8$, and so $\psi_s=g\cdot\phi_s$ and
$\psi_n=g\cdot\phi_n$. We have shown that the $9$ nilpotent elements
obtained from $9$ families in \cite[Table 7]{CD06} by setting
$a=b=c=d=0$ are pairwise inequivalent under the action of $\SL_8$,
hence the discussion in the paragraph above implies $\phi_n=\psi_n$.
Consequently, $\phi$ and $\psi$ belong to the same family.  The
families 12, 14, and 16 are ruled out because they have no
parameters. The family 1 is ruled out by Observation
\ref{pp:ClosedOrbits} since all $(\SL\rtimes S_4)$-orbits in this family
are closed orbits. The remaining families 2, 3, 6, 9, and 10 are ruled out
by the following Theorem \ref{lm:inv}.  Hence, our assumption must be
false, i.\,e., Theorem \ref{thm:main} is proved.

\begin{theorem}
\label{lm:inv}
Two states from
the same family in ~\cite[Table 7]{CD06} are in the same
$(\SL\rtimes S_4)$-orbit if and only if all invariants $f'\in \cB$
agree.
\end{theorem}

As a byproduct, Theorem~\ref{lm:inv} gives a complete answer to the
problem of SLOCC equivalence of pure $4$-qubit states, providing a
much simpler criterion for checking the SLOCC equivalence of pure
$4$-qubit states than the one proposed in \cite[Section 4]{CD06}.  To
prove this theorem, we need to examine all $9$ nine families. The case
of closed orbits has already been solved in \cite{CD06}, and the
assertion for the families 12, 14, and 16 holds trivially. The
detailed analysis of the families 2, 3, 6, 9, and 10 is given in
Appendix B.

\textit{The  LU case}---Having fully solved the SLOCC case, we now move
to the LU case. This is much harder, however we managed to deal with
the generic orbits, which in fact cover almost all orbits. Some facts
from our previous discussion of the SLOCC case will be used in the
proof.

Let
$f(a,b,c,d)=(a^2-b^2)(a^2-c^2)(a^2-d^2)(b^2-c^2)(b^2-d^2)(c^2-d^2)$, a
polynomial in four complex variables $a,b,c,d$. Furthermore, let
 \bea \notag
\Lambda &=& \{ g\cdot(ap_2+bp_4+cp_5-dp_6)\colon \\
&&\qquad g\in\SL,~ f(a,b,c,d)\ne0\}\sue W. \notag
 \eea

We observe that $\Lambda$ contains an open dense subset of $W$ which
is also $\SL\rtimes S_4$ invariant.  To prove it, we shall view $f$
as a polynomial function $f\colon\mathfrak{a}\to\bC$ by considering
$a,b,c,d$ as coordinates in $\mathfrak{a}$ with respect to the basis
$\{p_2,p_4,p_5,-p_6\}$. Then the polynomial $f^2$ extends (uniquely)
to an $g\in\cB$. Indeed, on $\mathfrak{a}$ we have
$27f^2=2^{15}(\Sigma^3-2\Pi^2)$, where $\Sigma$ and $\Pi$ are the
generators of $\cB$ of degree $8$ and $12$ from \cite{CD06}.  Recall that
the $4$-qubit hyperdeterminant, $\hdet$, is a homogeneous polynomial
$W\to\bC$ of degree $24$ which is $\SL\rtimes S_4$ invariant, i.e.,
$\hdet\in\cB$.  The set $\Omega=\{\psi\in W:\hdet(\psi)\ne0\}$ is
open, dense, and $\SL\rtimes S_4$ invariant subset of $W$.  It is known
\cite[Section III]{GW12} that each $\SL$-orbit, which is contained in
$\Omega$, meets $\mathfrak{a}$.  It follows that the set of all
$\psi\in\Omega$ such that $\Sigma(\psi)^3\ne2\Pi(\psi)^2$ is contained
in $\Lambda$, and clearly it is open and dense in $W$.

 \bt \label{thm:Unitary}
Let $\phi\in\Lambda$ and let $U\in\SU(8)$ be such that
$\psi:=U\cdot\phi\in W$. Then there exists $U'\in\SU\rtimes S_4$
such that $\psi=U'\cdot\phi$.
 \et

The proof of this theorem will be given in Appendix C.

\textit{Summary}---We have shown that the embedding of the space of
$4$-qubit pure states into the $4$-fermion space of $8$ modes
preserves the entanglement structure under the natural fermionic SLOCC
group $\SL_8$, which is also the case for generic orbits under the
fermionic LU group $\SU_8$. This surprising property of the $4$-qubit
states, following already known facts for $2$- and $3$-qubit systems,
reveals interesting connection between qubit and fermionic systems,
providing new perspectives on the entanglement structures of both systems.
One can naturally ask what happens for other LU orbits, and in the
more general case of $n$ qubits.
We believe that the discussion of these difficult, but intriguing
question shall shed light on insights of new physics in these many-body systems.

\textit{Acknowledgements}---LC was mainly supported by MITACS and
NSERC. The CQT is funded by the Singapore MoE and the NRF as part of
the Research Centres of Excellence programme. DD was supported in part
by an NSERC Discovery Grant. BZ is supported by NSERC and
CIFAR.


\appendix

\section{Appendix A: The algebras $\cA$ and $\cB$}\label{app:A}

This section discusses explicit generators of the algebras $\cA$ and
$\cB$ of polynomial invariants for the action of $\SL_8$ on $\we^4(V)$
and the action of $\SL\rtimes S_4$ on $\otimes_{i=1}^4 V_i$,
respectively, and their relationship.

The seven generators of $\cA$ have degrees $2$, $6$, $8$, $10$,
$12$, $14$, and $18$. They were computed by A. A. Katanova in
\cite{Katanova92}. Explicitly, they are given by the formulae \bea
\label{eq:Katanova} f_{2n}(\psi)=\tr A(\psi)^{2n},\quad
(n=1,3,4,5,6,7,9) \eea where $A(\psi)$ is a $28\times28$ matrix
whose entries are quadratic forms in the components of $\psi$. Once
we have chosen these generators, we obtain an explicit
identification of the variety $\we^4(V)/\SL_8$ with the affine space
$\bC^7$: given a closed $\SL_8$-orbit $\cO\sue\we^4(V)$ we chose a
point $\psi\in\cO$ and assign to $\cO$ the point in $\bC^7$ with
coordinates $f_{2n}(\psi)$, $n=1,3,4,5,6,7,9$.

The four generators of the algebra $\cB$ have degrees $2$, $6$, $8$,
and $12$.  They were computed first in \cite{CD06} and recently
another set of generators was computed in \cite{GW12}. Explicit
computation with these generators show that the restriction map
$\cA\to\cB$ induces an isomorphism of the subalgebra
$\bC[f_2,f_6,f_8,f_{12}]$ of $\cA$ onto $\cB$, i.\,e., we have
$\cB=\bC[f'_2,f'_6,f'_8,f'_{12}]$, where $f'_i$ denotes the
restriction of $f\in\cA$ to $\cB$. Consequently, the restrictions
$f'_{10}$, $f'_{14}$, $f'_{18}$ can be expressed as polynomials in
$f'_2$, $f'_6$, $f'_8$, $f'_{12}$. Explicitly we have obtained the
formulae
\begin{eqnarray}
2^9\cdot 3^4 f'_{10}&=& f'_2(7{f'_2}^4-2^5\cdot7\cdot9 f'_2f'_6
+2^6\cdot3^5 f'_8),
\end{eqnarray}
\begin{eqnarray}
2^{14}\cdot3^7\cdot5 f'_{14} &=& 2^5\cdot7\cdot11\cdot317 {f'_2}^4
f'_6 -11\cdot251 {f'_2}^7\nonumber\\
&&{}-2^{10}\cdot3^2\cdot7\cdot11\cdot13 f'_2{f'_6}^2 \nonumber\\
&&{}+2^{11}\cdot3^4\cdot7\cdot71 f'_2 f'_{12}\nonumber\\
&&{}+2^{11}\cdot3^5\cdot7\cdot11 f'_6 f'_8\nonumber\\
&&{}-2^6\cdot3^2\cdot7\cdot11\cdot103 {f'_2}^3 f'_8,
\end{eqnarray}
\begin{eqnarray}
2^{19}\cdot3^9\cdot5^2 f'_{18} &=& -5^2\cdot13903{f'_2}^9\nonumber\\
&&{}+2^7\cdot 5\cdot89\cdot1609{f'_2}^6{f'_6}\nonumber\\
&&{}-2^7\cdot3^2\cdot5\cdot8989{f'_2}^5{f'_8}\nonumber\\
&&{}+2^{12}\cdot3^2\cdot37\cdot109{f'_2}^3{f'_6}^2 \nonumber\\
&&{}+2^{10}\cdot5^2\cdot7^2\cdot13513{f'_2}^3{f'_{12}}\nonumber\\
&&{}-2^{15}\cdot3^6\cdot349{f'_2}^2{f'_6}{f'_8}  \nonumber\\
&&{}+2^{12}\cdot3^9\cdot331{f'_2}{f'_8}^2\nonumber\\
&&{}-2^{21}\cdot3^5\cdot5{f'_6}^3 \nonumber\\
&&{}+2^{12}\cdot5\cdot71\cdot127\cdot1409{f'_6}{f'_{12}}.
\end{eqnarray}

The image of the morphism $\Phi$ can be described as the graph of
the morphism $\bC^4\to\bC^3$ given by the above three equations.
More precisely, we have to substitute $f'_2,f'_6,f'_8,f'_{12}$ with
complex coordinates $z_1,z_2,z_3,z_4$ and $f'_{10},f'_{14},f'_{18}$
with $z_5,z_6,z_7$, respectively, to obtain the formulae expressing
$z_5,z_6,z_7$ as polynomial functions in $z_1,z_2,z_3,z_4$.

\section{Appendix B: Dynkin diagrams for 
$(\mathfrak{so}_8,\mathfrak{sl}_2^{\times4})\sue
(\mathfrak{e}_7,\mathfrak{sl}_8)$}

In this section we describe the embedding of the symmetric space 
$(\mathfrak{so}_8,\mathfrak{sl}_2^{\times4})$ into the larger symmetric space $(\mathfrak{e}_7,\mathfrak{sl}_8)$ by using the root system and the root space decomposition of $\mathfrak{e}_7$. 

The simple roots of $\mathfrak{e}_7$ are 
$\alpha_1,\alpha_2,\ldots,\alpha_7$. The simple roots of the subalgebra $\mathfrak{sl}_8$ are 
$-\alpha_0,\alpha_1,\alpha_3,\alpha_4,\ldots,\alpha_7$, where 
$\a_0=2\a_1+2\a_2+3\a_3+4\a_4+3\a_5+2\a_6+\a_7$ is the highest root of $\mathfrak{e}_7$.
These seven roots are painted in white to indicate that the corresponding root vectors belong to $\mathfrak{sl}_8$, the $+1$ eigenspace of the involution $\theta$. 
The roots $\alpha_2$ and $\beta$ are painted in black because the corresponding root vectors belong to the subspace $\mathfrak{p}=\wedge^4(V)$, the $-1$ eigenspace of $\theta$.

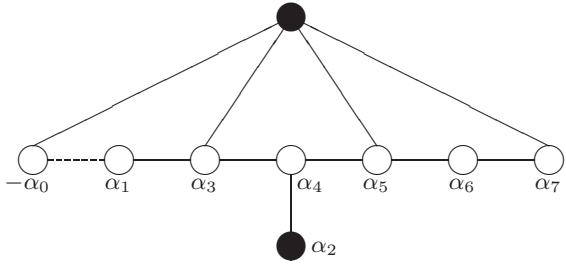
\begin{figure}[h]
\setlength{\unitlength}{0.15in} 
\centering 
\begin{picture}(26,9) 
\put(5,9){$\beta=\a_1+\a_2+\a_3+2\a_4+\a_5+\a_6$}

\put(12,8){\circle*{1}}

\put(3,3){\circle{1}}
\put(6,3){\circle{1}}
\put(9,3){\circle{1}}
\put(12,3){\circle{1}}
\put(15,3){\circle{1}}
\put(18,3){\circle{1}}
\put(21,3){\circle{1}}

\put(12,0){\circle*{1}}

\put(12,2.5){\line(0,-1){3}}

\put(3.5,3){\line(1,0){0.2}}
\put(3.8,3){\line(1,0){0.2}}
\put(4.1,3){\line(1,0){0.2}}
\put(4.4,3){\line(1,0){0.2}}
\put(4.7,3){\line(1,0){0.2}}
\put(5.0,3){\line(1,0){0.2}}
\put(5.3,3){\line(1,0){0.2}}

\put(6.5,3){\line(1,0){2}}
\put(9.5,3){\line(1,0){2}}
\put(12.5,3){\line(1,0){2}}
\put(15.5,3){\line(1,0){2}}
\put(18.5,3){\line(1,0){2}}

\put(3,3.5){\line(2,1){9}}
\put(9,3.5){\line(2,3){3}}
\put(15,3.5){\line(-2,3){3}}
\put(21,3.5){\line(-2,1){9}}

\put(2,2){$-\alpha_0$}
\put(5.5,2){$\alpha_1$}
\put(8.5,2){$\alpha_3$}
\put(12.2,2){$\alpha_4$}
\put(14.5,2){$\alpha_5$}
\put(17.5,2){$\alpha_6$}
\put(20.5,2){$\alpha_7$}

\put(12.7,-0.2){$\alpha_2$}

\end{picture}
\caption{Embedding of $\mathfrak{so}_8$ into $\mathfrak{e}_7$}
\label{fig:digram}
\end{figure}

The roots $\a_3,\beta,\a_5,\a_7$ form the Dynkin diagram of the 
subalgebra $\mathfrak{so}_8$. It is interesting that $\a_0$ is also the highest root of this $\mathfrak{so}_8$. The intersection $\mathfrak{so}_8\cap\mathfrak{sl}_8$ is the Lie algebra 
$\mathfrak{sl}_2^{\times4}$ of $\SL$. The simple roots of this subalgebra are $-\a_0$, $\a_3$, $\a_5$ and $\a_7$. 

Moreover, $\a_2$ is the lowest weight of the $\SL_8$ module 
$\wedge^4(V)$, and $\beta=(-\a_0+\a_3+\a_5+\a_7)/2$ is the lowest weight of the $\SL$ module $\otimes_{i=1}^4 V_i$. 

We remark that the point-wise stabilizer of $\mathfrak{c}$ in $\SL_8$, i.\,e., the group
\begin{alignat}{5}
A=\{ a\in\SL_8\colon a\cdot p_i=p_i\;\forall i=1,\ldots,7\}
\end{alignat}
is the three-qubit Pauli group of order $256$. The centre $Z(A)$ of
$A$ is generated by $i I_8$, where $i^2=-1$ and $I_8$ is the identity
of $\SL_8$. Hence the action of $A$ on $\wedge^4 V$ is the abstract
group $A/Z(A)$ which is an elementary Abelian group of order
$2^6$. This agrees with \cite[Summary Table p. 261, No. 18]{PV94}.

\section{Appendix C: Proof of Theorem~\ref{lm:inv}}\label{app:B}

This section proves Theorem~\ref{lm:inv}. We explicitly list the
orbits from \cite[Table 7]{CD06} with non-trivial Jordan
decomposition as in Table~\ref{tab:mixed_orbits}.

\begin{table}[tbh]
\def\arraystretch{1.5}
\def\s{\rule{0pt}{\baselineskip}}
\begin{tabular}{c|p{0.4\textwidth}}
no. & representative $|\psi\rangle$\\
\hline
2 & $\frac{a+c-i}{2}(|0000\rangle+|1111\rangle)+\frac{a-c+i}{2}(|0011\rangle+|1100\rangle)$\newline$\s+\frac{b+c+i}{2}(|0101\rangle+|1010\rangle)+\frac{b-c-i}{2}(|0110\rangle+|1001\rangle)$\newline$\s+\frac{i}{2}(|0001\rangle+|0111\rangle+|1000\rangle+|1110\rangle$\newline$\s  -|0010\rangle-|0100\rangle-|1011\rangle-|1101\rangle)$\\
\hline
3 & $\frac{a}{2}(|0000\rangle+|1111\rangle +|0011\rangle+1100\rangle)$\newline$\s + \frac{b+1}{2}(|0101\rangle+|1010\rangle)  +\frac{b-1}{2}(|0110\rangle+|1001\rangle)$\newline$\s+\frac{1}{2}(|1101\rangle+|0010\rangle-|0001\rangle-|1110\rangle)$\\
\hline
6 & $\frac{a+b}{2}(|0000\rangle+|1111\rangle)+b(|0101\rangle+|1010\rangle)
$\newline$\s+i(|1001\rangle-|0110\rangle)  +\frac{a-b}{2}(|0011\rangle+|1100\rangle)$\newline$\s+\frac{1}{2}(|0010\rangle+|0100\rangle+|1011\rangle+|1101\rangle
$\newline$\s -|0001\rangle-|0111\rangle-|1000\rangle-|1110\rangle)$\\
\hline
9 & $a(|0000\rangle+|0101\rangle+|1010\rangle+|1111\rangle)
$\newline$\s -2i(|0100\rangle-|1001\rangle-|1110\rangle)$\\
\hline
10 & $\frac{a+i}{2}(|0000\rangle+|1111\rangle
+|0011\rangle+1100\rangle)
$\newline$\s+\frac{a-i+1}{2}(|0101\rangle+|1010\rangle)
+\frac{a-i-1}{2}(|0110\rangle+|1001\rangle)$\newline$\s+\frac{i+1}{2}(|1101\rangle+|0010\rangle)
+\frac{i-1}{2}(|0001\rangle+|1110\rangle)
$\newline$\s-\frac{i}{2}(|0100\rangle+|0111\rangle+|1000\rangle+|1011\rangle)$
\end{tabular}
\caption{Orbits from \cite[Table 7]{CD06} with non-trivial Jordan
decomposition.
\label{tab:mixed_orbits}}
\end{table}

As already discussed, it suffices to show that
\textit{two states $|\psi(a,b,c)\rangle$ and $|\psi(a',b',c')\rangle$ from
the same family in Table \ref{tab:mixed_orbits} are in the same
$(\SL\rtimes S_4)$-orbit if and only if all invariants $f'\in \cB$
agree.}

If the invariants do not agree, then the states are obviously in
different orbits. In order to show sufficiency, assume that the
invariants agree, i.\,e., $g_j(a,b,c)=g_j(a',b',c')$ for $j=2,6,8,12$,
where $g_j(a,b,c)=f'_j(|\psi(a,b,c)\rangle)$ and the polynomials
$f'_j$ are the generators of the algebra $\cB$. For each family, we
obtain a system of polynomial equations.  The corresponding radical
ideal is generated by the polynomials listed in Table
\ref{tab:radical_generators}. Computing the primary decomposition of
the ideals, we find that there are linear relations between the
triples of variables $(a,b,c)$ and $(a',b',c')$ given by finite
groups (see Table \ref{tab:variety_symmetry}).

\begin{table}[h!]
\centerline{
\def\arraystretch{1}
\def\s{\rule[-1ex]{0pt}{\baselineskip}}
\begin{tabular}{c|p{0.4\textwidth}}
no. & generators of the radical ideal\\
\hline 2 &
$(c-c')(c+c')(c-a'/2-b'/2)(c-a'/2+b'/2)$\newline${}\quad\times(c+a'/2-b'/2)(c+a'/2+b'/2)$\s\newline
$b'^4+2b'^2c^2-3c^4-b'^2a'^2+c^2a'^2-b'^2b'^2+c^2b'^2$\newline${}\quad+a'^2b'^2-2b'^2c'^2+2c^2c'^2-a'^2c'^2-b'^2c'^2+c'^4$\s\newline
$a^2+b'^2+2c^2-a'^2-b'^2-2c'^2$\\
\hline
3 &$(b-b')(b+b')(b-a')(b+a')$\s\newline
$a^2+b^2-a'^2-b'^2$\\
\hline
6 &
$(b-b')(b+b')(b-a'/2-b'/2)(b-a'/2+b'/2)$\newline${}\quad\times(b+a'/2-b'/2)(b+a'/2+b'/2)$\s\newline
$a^2+3b^2-a'^2-3b'^2$\\
\hline
9 & $(a-a')(a+a')$\\
\hline
10 & $(a-a')(a+a')$
\end{tabular}}
\caption{Generators of the radical of the ideal generated by
  $g_j(a',b',c')-g_j(a,b,c)$.\label{tab:radical_generators}}
\end{table}

\begin{table}
\centerline{
\def\arraystretch{1.25}
\def\s{\rule{0pt}{\baselineskip}}
\begin{tabular}{c|c|c|c}
no. & group & order & generators\\
\hline
2 & $S_4\times Z_2$ & 48 & $(a',b',c')=(b,-a,c)$\\
  &                 &    & $(a',b',c')$\\
   &                 &    & $((a+b)/2+c,(a+b)/2-c,(a-b)/2)$\\
\hline
3 & $D_4$           &  8 & $(a',b')=(a,-b)$\\
  &                 &    & $(a',b')=(b,a)$\\
\hline
6 & $D_6$           & 12 & $(a',b')=(a+3b,a-b)/2$\\
  &                 &    & $(a',b')=(a,-b)$\\
\hline
9 & $Z_2$           &  2 & $a'=-a$\\
\hline 10& $Z_2$           &  2 & $a'=-a$
\end{tabular}}
\caption{Symmetries of the varieties corresponding to
  identical invariants.\label{tab:variety_symmetry}}
\end{table}

Assume that for the states $|\psi^{(\mu)}(a,b,c)\rangle$ and
$|\psi^{(\mu)}(a',b',c')\rangle$ from the same family
$\mu=2,3,6,9,10$ all polynomial invariants agree. In the following,
we show that the linear transformations on the variables $a,b,c$
corresponding to the generators of the groups in Table
\ref{tab:variety_symmetry} can be realized by operations from
$\SL\rtimes S_4$ on the states.

\begin{description}
\item[Family 2] Direct computation shows that (i) applying the
  transformation
  \begin{eqnarray}
  &&\frac{1}{\sqrt{2}}\begin{pmatrix} 1 & -i\\-i &  1\end{pmatrix}\otimes
  \frac{1}{\sqrt{2}}\begin{pmatrix} 1 & -i\\-i &  1\end{pmatrix}\nonumber\\
  &&\quad\otimes
  \frac{1}{\sqrt{2}}\begin{pmatrix} i &  1\\-1 & -i\end{pmatrix}\otimes
  \frac{1}{\sqrt{2}}\begin{pmatrix}-i & -1\\ 1 &  i\end{pmatrix}
  \end{eqnarray}
  followed by a permutation of the last two qubits maps
  $|\psi^{(2)}(a,b,c)\rangle$ to $|\psi^{(2)}(b,-a,c)\rangle$; (ii)
  swapping the two middle qubits maps the state $|\psi^{(2)}(a,b,c)\rangle$ to the state
  $|\psi^{(2)}((a+b)/2+c,(a+b)/2-c,(a-b)/2)\rangle$.
\item[Family 3] Direct computations shows that
 (i)  the states $|\psi^{(3)}(a,b)\rangle$ and
    $|\psi^{(3)}(a,-b)\rangle$ are related by the transformation
    $I_2\otimes I_2\otimes(i\sigma_y)\otimes(i\sigma_y)$;
 (ii) the states $|\psi^{(3)}(a,b)\rangle$ and
    $|\psi^{(3)}(b,a)\rangle$ are related by the transformation
    $M^{-1}\otimes M^{-1}\otimes M\otimes M$, where
    \begin{equation}\label{eq:matrix_M}
    M=\frac{1}{\sqrt{2}}
    \begin{pmatrix}
    1 & i
    \\i & 1
    \end{pmatrix}.
    \end{equation}
\item[Family 6] Direct computation shows that swapping the two middle
  qubits maps the state $|\psi^{(6)}(a,b)\rangle$ to the state
  $|\psi^{(6)}((a+3b)/2,(a-b)/2)\rangle$.  Furthermore, the following
  calculation shows that applying the transformation $T_1=I_2\otimes
  I_2\otimes(i\sigma_y)\otimes(i\sigma_y)$, followed by swapping the
  first two qubits maps the state $|\psi^{(6)}(a,b)\rangle$ to the
  state $|\psi^{(6)}(a,-b)\rangle$:
\begin{eqnarray}
\noalign{$|\psi{}^{(6)}(a,b)\rangle$}
\;&{}=&
\frac{a+b}{2}(|0000\rangle+|1111\rangle)+b(|0101\rangle+|1010\rangle)\nonumber\\
&&{}+i(|1001\rangle-|0110\rangle)+\frac{a-b}{2}(|0011\rangle+|1100\rangle)\nonumber\\
&&{}+\frac{1}{2}(|0010\rangle+|0100\rangle+|1011\rangle+|1101\rangle\nonumber\\
&&{}-|0001\rangle-|0111\rangle-|1000\rangle-|1110\rangle)\nonumber\\
&&\llap{$\stackrel{T_1}{\longrightarrow}\;$}
\frac{a+b}{2}(|0011\rangle+|1100\rangle)-b(|0110\rangle+|1001\rangle)\nonumber\\
&&{}-i(|1010\rangle-|0101\rangle)+\frac{a-b}{2}(|0000\rangle+|1111\rangle)\nonumber\\
&&{}+\frac{1}{2}(-|0001\rangle+|0111\rangle+|1000\rangle-|1110\rangle\nonumber\\
&&{}+|0010\rangle-|0100\rangle-|1011\rangle+|1101\rangle)\nonumber\\
&&\llap{$\stackrel{\tau=(1\,2)}{\longrightarrow}\;$}
\frac{a+b}{2}(|0011\rangle+|1100\rangle)-b(|1010\rangle+|0101\rangle)\nonumber\\
&&{}-i(|0110\rangle-|1001\rangle)+\frac{a-b}{2}(|0000\rangle+|1111\rangle)\nonumber\\
&&{}+\frac{1}{2}(-|0001\rangle+|1011\rangle+|0100\rangle-|1110\rangle\nonumber\\
&&{}+|0010\rangle-|1000\rangle-|0111\rangle+|1101\rangle)\nonumber\\
&=&|\psi{}^{(6)}(a,-b)\rangle
\end{eqnarray}
\item[Family 9] Direct computation shows that the states
  $|\psi^{(9)}(a)\rangle$ and $|\psi^{(9)}(-a)\rangle$ are related by the transformation
  $I_2\otimes(i\sigma_z)\otimes I_2\otimes(i\sigma_z)$.
\item[Family 10] Direct computation shows that the states
  $|\psi^{(10)}(a)\rangle$ and $|\psi^{(10)}(-a)\rangle$ are related by the transformation
  $M^{\otimes 4}$ where $M$ is given in (\ref{eq:matrix_M}).
\end{description}
In summary, we have shown that two states which belong to the same
family and for which all polynomial invariants agree are in the same
$(\SL\rtimes S_4)$-orbit. The finite groups in Table
\ref{tab:variety_symmetry} define relations on the space of
parameters $(a,b,c)$.

\section{Appendix D: Proof of Theorem \ref{thm:Unitary}}\label{app:C}

In this section we prove Theorem \ref{thm:Unitary}.

By the hypothesis we have $\phi=g\cdot\alpha$ for some $g\in\SL$ and
some $\alpha=ap_2+bp_4+cp_5-dp_6$, where $a,b,c,d\in\bC$ and
$a^2,b^2,c^2,d^2$ are pairwise distinct. After setting
$v_i=g\cdot\ket{i}$, $i=1,\ldots,8$, we have
 \bea
\phi
 &=& a(v_1\wedge v_3\wedge v_5\wedge v_7
      +v_2\wedge v_4\wedge v_6\wedge v_8) \notag \\
 &+& b(v_1\wedge v_3\wedge v_6\wedge v_8
      +v_2\wedge v_4\wedge v_5\wedge v_7) \notag \\
 &+& c(v_1\wedge v_4\wedge v_5\wedge v_8
      +v_2\wedge v_3\wedge v_6\wedge v_7) \notag \\
 &+& d(v_1\wedge v_4\wedge v_6\wedge v_7
      +v_2\wedge v_3\wedge v_5\wedge v_8).\notag
 \eea
We set $u_i=U^\dag\ket{i}$ for $i=1,\ldots,8$, and so
$\{u_i\}$ is an orthonormal basis of $V$. We have
 \bea \notag
v_j &=& \sum_{i=1}^8 x_{ij} u_i, \quad j=1,\ldots,8
 \eea
where $x_{ij}=\braket{u_i}{v_j}=\bra{i}Ug\ket{j}$. Thus
$X:=(x_{ij})$ equals the matrix $Ug$, and we partition it into $16$
blocks $X_{kl}$ of size $2\times2$.

For convenience, set $e_{ij}=\ket{i\wedge j}$.
Since $\psi=U\cdot\phi\in W$, the partial inner product
$\braket{e_{2k-1,2k}}{\psi}$ vanishes for $k=1,2,3,4$.
Equivalently, we have
 \bea \label{eq:parc}
\braket{u_{2k-1}\wedge u_{2k}}{\phi}=0, \quad k=1,2,3,4.
 \eea

Let us consider this equation for $k=1$. After expanding the partial
inner product by using the formula given in \cite[Eq. (2)]{CCD+13}, we
obtain a linear combination of the bivectors $v_i\wedge v_j$ with
$1\le i<j\le8$. The bivectors $v_{2k-1}\wedge v_{2k}$ do not occur in
this expansion. The coefficients of the other $24$ bivectors
$v_i\wedge v_j$ must be $0$, and so we obtain $24$ equations. Each pair
of the parameters $a,b,c,d$ occurs in exactly four of these equations.
For instance, the four equations in which only $a$ and $b$ occur are
the following:
\bea
v_1\wedge v_3\colon &&\quad \notag aD_{57}+bD_{68}=0, \\
v_2\wedge v_4\colon &&\quad \notag bD_{57}+aD_{68}=0, \\
v_5\wedge v_7\colon &&\quad \notag aD_{13}+bD_{24}=0, \\
v_6\wedge v_8\colon &&\quad \notag bD_{13}+aD_{24}=0,
\eea
where
$D_{ij}=x_{1i}x_{2j}-x_{1j}x_{2i}$.  Since $a^2\ne b^2$, the first two
equations imply that $D_{57}=D_{68}=0$ and the last two imply that
$D_{13}=D_{24}=0$.  One obtains similar results by using the other
five pairs of the parameters $a,b,c,d$. The final result is that all
$D_{ij}$, $i<j$, vanish except possibly $D_{12}$, $D_{34}$, $D_{56}$,
and $D_{78}$. As $X$ is invertible, at least one of these four minors
does not vanish. It is now easy to see that exactly one of the blocks
$X_{1l}$ is invertible, and all others vanish.

By applying the same arguments to the other three equations in
\eqref{eq:parc}, we deduce that in each row and each column of blocks
in $X$ exactly one block is invertible and all others vanish. Since
$X=Ug$ and $g\in\SL$ is block-diagonal, it follows that the unitary
matrix $U=Xg^{-1}$ has a permuted block structure, i.e.,
$U\in\Un(2)^{\times4}\rtimes S_4$.  For suitable $S=\oplus_{k=1}^4
\lambda_k I_2$, with $\prod\lambda_k=1$, we have $US\in\SU\rtimes
S_4$. By using the facts that $Sg=gS$ and $S\cdot w=w$ for all $w\in
W$, we obtain that
$US\cdot\phi=USg\cdot\alpha=Ug\cdot\alpha=\psi$. Thus we can take
$U'=US$ to complete the proof.

\end{document}